%% file: DLforAQP.tex
\documentclass[conference]{IEEEtran}

\usepackage{subfigure}
\usepackage{balance}
\usepackage{graphicx}
\usepackage{amsmath}
\usepackage{amssymb}
\usepackage{algorithm}
\usepackage[noend]{algorithmic}
\usepackage{multirow}
\usepackage{url}
\usepackage{xcolor}
\usepackage{soul}
\usepackage{xspace}

\usepackage{cleveref}

\newcommand{\stitle}[1]{\vspace{1ex}\noindent{\bf #1}}

\newcommand{\add}[1]{{#1}}

\newcommand{\ie}{{\em i.e.,}\xspace}
\newcommand{\eg}{{\em e.g.,}\xspace}

\title{
Approximate Query Processing for Data Exploration using Deep Generative Models\\
}

\author
 {
 \IEEEauthorblockN{
     Saravanan Thirumuruganathan$^{\ddag\dag}$,
     Shohedul Hasan$^{\ddag}$,
     Nick Koudas$^{\ddag\ddag}$,
     Gautam Das$^{\ddag}$
     }

 \IEEEauthorblockA{
     $^{\ddag\dag}$QCRI, HBKU;
     $^{\ddag}$University of Texas at Arlington;
     $^{\ddag \ddag}$University of Toronto}

     	$^{\ddag\dag}$sthirumuruganathan@hbku.edu.qa,
      	$^{\ddag}$\{shohedul.hasan@mavs,~gdas@cse\}.uta.edu,
     	$^{\ddag\ddag}$koudas@cs.toronto.edu

 }

\begin{document}

\maketitle

\input{abstract}

\maketitle

\input{introduction}
\input{preliminaries}
\input{background}
\input{VAEforAQP}

\input{modelEnsembles}
\input{experiments}
\input{relatedWork}
\input{conclusion}

\section{Acknowledgments}
The work of Gautam Das was supported in part by grants 1745925 and 1937143 from the National Science Foundation, and a grant from AT\&T.

\balance
\interlinepenalty=10000
\bibliographystyle{abbrv}
\bibliography{DLforAQP}

\end{document}

%% file: abstract.tex
\begin{abstract}
Data is generated at an unprecedented rate surpassing our ability to analyze them.
The database community has pioneered many novel techniques for Approximate Query Processing (AQP)
that could give approximate results in a fraction of time needed for computing exact results.
In this work, we explore the usage of deep learning (DL) for answering
aggregate queries specifically for interactive applications such as data exploration and visualization.
We use \emph{deep generative models}, an unsupervised learning based approach,
to learn the data distribution faithfully such that
aggregate queries could be answered approximately by \emph{generating} samples from the learned model.
The model is often compact --  few hundred KBs --
so that arbitrary AQP queries could be answered on the client side without contacting the database server.
Our other contributions include
identifying model bias and minimizing it through a rejection sampling based approach and
an algorithm to build model ensembles for AQP for improved accuracy.
Our extensive experiments show that our proposed approach can provide answers with high accuracy and low latency.
\end{abstract}

%% file: introduction.tex
\section{Introduction}
\label{sec:introduction}

Data driven decision making has become the dominant paradigm for businesses seeking to gain an edge over competitors.
However, the unprecedented rate at which data is generated surpasses our ability to analyze them.
Approximate Query Processing (AQP) is a promising technique that provides \emph{approximate} answers to
queries at a fraction of the cost needed to answer it exactly.
AQP has numerous applications in data exploration and visualization
where approximate results are acceptable as long as they can be obtained near real-time.

\stitle{Case Study.}
Consider an user who performs data exploration and visualization on a popular dataset such as NYC Taxi dataset.
The user issues ad-hoc aggregate queries, involving arbitrary subsets of attributes of interest, such as \emph{what is the average number of passengers on trips starting from Manhattan?}
or \emph{what is the average trip duration grouped by hour?} and so on.
Since this is for exploratory purposes, an imprecise answer is often adequate.
A traditional approach is to issue aggregate queries to the database server,
get exact or approximate answers accessing the base data or pre-computed/on-demand samples and display the returned results to the user.
However, this could suffer from high latency that is not conducive for interactive analysis.
In this paper, we propose an alternate approach where the approximate results could be computed
entirely at the client side.
Specifically, we build a deep generative model that approximates the data distribution with high fidelity and is lightweight (few hundreds of KBs).
This model is sent to the client and could be used to generate \emph{synthetic} samples
over which AQP could be performed locally on arbitrary subsets of attributes, without any communication with the server.
Our approach is complementary to traditional AQP exploring a new research direction of utilizing deep generative models for data exploration. It offers a lightweight model that can answer arbitrary queries which we experimentally demonstrate exhibit superior accuracy. For queries requiring provable guarantees we default to traditional AQP or exact query evaluation.

\subsection{Outline of Technical Results}

\stitle{Deep Learning for AQP.}
Deep Learning (DL) \cite{Goodfellow-et-al-2016} has become popular due to its excellent performance in many complex applications.
In this paper, we investigate the feasibility of using DL for answering aggregate queries for data exploration and visualization.
Structured databases seem intrinsically different from prior areas where DL has shined - such as computer vision and natural language processing.
Furthermore, the task of generating approximate estimates for an aggregate query is quite different from common DL tasks.
However, we show that AQP can be achieved in an effective and efficient manner using DL models.


\stitle{Deep Generative Models for AQP.}
Our key insight is to train a DL model to learn the data distribution of the underlying data set effectively.
Once such a model is trained, it acts as a concise representation of the dataset.
The samples generated from the model have a data distribution that is almost identical to that of the underlying dataset.
Hence, existing AQP techniques~\cite{mozafari2015handbook,Mozafari:2017:AQE:3035918.3056098} could be transparently applied on these samples.
Furthermore, the model could generate as many samples as required without the need to access the underlying dataset.
This makes it very useful for interactive applications as all the computations could be done locally.

\stitle{Technical Challenges.}
The key challenge is to identify a DL based distribution estimation approach that is
expressive enough to reflect statistical properties of real-world datasets and yet tractable and efficient to train.
It must be non-parametric and not make any prior assumption about data characteristics.
A large class of DL techniques - dubbed collectively as deep generative models - could be used for this purpose.
Intuitively, a deep generative model is an unsupervised approach that learns
the probability distribution of the dataset from a set of tuples.
Often, learning the exact distribution is challenging, thus generative models
learn a model that is very similar to the true distribution of the underlying data.
This is often achieved through neural networks that learn a function that maps the approximate distribution to the true distribution.
Each of the generative models have their respective advantages and disadvantages.
We focus on variational autoencoders~\cite{doersch2016tutorial} that aim to learn a low dimensional latent representation of the training data that optimizes the log-likelihood of the data through evidence lower bound.

%% file: preliminaries.tex
\section{Preliminaries}
\label{sec:preliminaries}

Consider a relation $R$ with $n$ tuples and $m$ attributes $A_1, A_2, \ldots, A_m$.
Given a tuple $t$ and an attribute $A_i$, we denote the value of $A_i$ in $t$ as $t[A_i]$.
Let $Dom(A_i)$ be the domain of attribute $A_i$.

\stitle{Queries for AQP.}
In this paper, we focus on aggregate analytic queries of the general format:~\\

\texttt{\null \qquad SELECT g, AGG(A) FROM R\\
	\null \qquad ~ WHERE filter GROUP BY G\\
}

Of course, both the WHERE and GROUP BY clauses are optional.
Each attribute $A_i$ could be used as a \emph{filter} attribute involved in a predicate or
as a \emph{measure} attribute involved in an aggregate.
The filter could be a conjunctive or disjunctive combination of conditions.
Each of the conditions could be any relational expression of the format \texttt{A op CONST}
where $A$ is an attribute and $op$ is one of $\{=, \neq, <, >, \leq, \geq\}$.
AGG could be one of the standard aggregates \texttt{AVG, SUM,  COUNT} that have been extensively studied in prior AQP literature.
One could use other aggregates such as \texttt{QUANTILES} as long as a statistical estimator exists to generate aggregate estimates.

\stitle{Performance Measures.}
Let $q$ be an aggregate query whose true value is $\theta$.
Let $\tilde{\theta}$ be the estimate provided by the AQP system.
Then, we can measure the estimation accuracy through relative error defined as
\begin{equation}
	RelErr(q) = \frac{|\tilde{\theta} - \theta|}{\theta}
	\label{eq:relErr}
\end{equation}

For a set of queries $Q = \{q_1, \ldots, q_r\}$, the effectiveness of the AQP system could be computed through average relative error.
Let $\theta_j$ and $\tilde{\theta_j}$ be the true and estimated value of the aggregate for query $q_j$.
\begin{equation}
	AvgRelErr(Q) = \frac{1}{r} \sum_{j=1}^{r} \frac{|\tilde{\theta_j} - \theta_j|}{\theta_j}
	\label{eq:avgRelErr}
\end{equation}

We could also use the average relative error to measure the accuracy of the estimate for \texttt{GROUP BY} queries.
Suppose we are given a group by query $q$ with groups $G=\{g_1, \ldots, g_r\}$.
It is possible that the sample does not contain all of these groups
and the AQP system generates estimates for groups $\{g_{j_1}, \ldots, g_{j_r'}\}$ where each $g_{j_i} \in G$.
As before, let $\theta_{j_i}$ and $\widetilde{\theta_{j_i}}$ be the true and estimated value of the aggregate for group $g_{j_i}$.
By assigning 100\% relative error for missing groups, the average relative error for group by queries is defined as,
\begin{equation}
	AvgRelErr(q) = \frac{1}{r} \left( (r - r') +  \sum_{i=1}^{r'} \frac{|\widetilde{\theta_{j_i}} - \theta_{j_i}|}{\theta_{j_i}} \right)
	\label{eq:avgRelErrGp}
\end{equation}

%% file: background.tex
\section{Background}
\label{sec:background}
In this section, we provide necessary background about generative models and variational autoencoders in particular.

\stitle{Generative Models.}
Suppose we are given a set of data points $X=\{x_1, \ldots, x_n\}$ that are distributed according to some unknown probability distribution $P(X)$.
Generative models seek to learn an approximate probability distribution $Q$ such that $Q$ is very similar to $P$.
Most generative models also allow one to generate samples $X'=\{x'_1, \ldots, \}$ from the model $Q$
such that the $X'$ has similar statistical properties to $X$.
Deep generative models use powerful function approximators (typically, deep neural networks) for learning to approximate the distribution.

\stitle{Variational Autoencoders (VAEs).}
VAEs are a class of generative models~\cite{doersch2016tutorial,bengio2015deep,jaanVAE}
that can model various complicated data distributions and generate samples.
They are very efficient to train, have an interpretable latent space and could be adapted effectively to different domains such as images, text and music.
%
Latent variables are an intermediate data representation that captures {\em data characteristics} used for generative modelling.
Let $X$ be the relational data that we wish to model and $z$ a latent variable.
\add{Let $P(X)$ be the probability distribution from which the underlying relation
consisting of attributes $A_1, \ldots, A_m$ was derived
and $P(z)$ as the probability distribution of the latent variable.}
Then $P(X|z)$ is the distribution of generating data given latent variable.
We can model $P(X)$ in relation to $z$ as $P(X)=\int P(X|z)P(z)dz$ marginalizing $z$ out of the joint probability $P(X,z)$. The challenge is that we do not know $P(z)$ and $P(X|z)$. The underlying idea in variational modelling is to infer $P(z)$ using $P(z|X)$.

\stitle{Variational Inference.}
We use a method called Variational Inference (VI) to infer $P(z|X)$ in VAE.
The main idea of VI is to approach inference  as an optimization problem. We model the true distribution $P(z|X)$ using a simpler distribution (denoted as $Q$) that is easy to evaluate, e.g. Gaussian, and minimize the difference between those two distribution using KL divergence metric, which tells us how different $P$ is from  $Q$.
\add{Typically, the simpler distribution depends on the attribute type.
Gaussian distribution is often appropriate for real numbers
while Bernoulli distribution is often used for categorical attributes.}
Assume we wish to infer $P(z|X)$ using $Q(z|X)$. The KL divergence is specified as:
\begin{equation}
\begin{split}
D_{KL}[Q(z|X)||P(z|X)]=\sum_{z}Q(z|X)\log(\frac{Q(z|X)}{P(z|X)}) = \\
E[\log(\frac{Q(z|X)}{P(z|X)})]=E[\log(Q(z|X)) - \log(P(z|X))]
\end{split}
\end{equation}
We can connect~\cite{doersch2016tutorial} $Q(z|X)$ which is a projection
of the data into the latent space and $P(X|z)$ which generates data given a latent variable $z$ through
Equation \ref{eq:var2} that is also called as the variational objective.
\begin{equation}
\label{eq:var2}
\begin{split}
    \log P(X)-D_{KL}[Q(z|X||P(z|X)] \\ =E[\log P(X|z)]-D_{KL}[Q(z|X||P(z)]
    \end{split}
\end{equation}

\stitle{Encoders and Decoders.}
A different way to think of this equation is
as $Q(z|X)$ encoding the data using $z$ as an intermediate data representation and $P(X|z)$ generates data given a latent variable $z$. Typically
$Q(z|X)$ is implemented with a neural network mapping the underlying data space into the latent space ({\em encoder network}).
Similarly $P(X|z)$ is implemented
with a neural network and is responsible to generate data following the distribution $P(X)$ given sample latent variables $z$ from the latent space ({\em decoder network}).
The variational objective has a very natural interpretation. We wish to model our data $P(X)$ under some error function $D_{KL}[Q(z|X||P(z|X)]$.
In other words, VAE tries to identify the lower bound of $\log(P(X))$, which in practice is good enough as trying to determine the exact distribution is often intractable. For this we aim to maximize over some mapping from latent variables to $\log P(X|z)$ and minimize the difference between our simple distribution $Q(z|X)$ and the true latent distribution $P(z)$. Since we need to sample from $P(z)$ in VAE typically one chooses a simple distribution to sample from such as $N(0,1)$. Since we wish to minimize the distance between $Q(z|X)$ and $P(z)$ in VAE one typically assumes that $Q(z|X)$ is also normal with mean $\mu(X)$ and variance $\Sigma(X)$.
Both the encoder and the decoder networks are trained end-to-end. After training, data can be generated by sampling $z$ from a normal distribution and passing it to the decoder network.


%% file: VAEforAQP.tex
\section{AQP Using Variational AutoEncoders}
\label{sec:VAEforAQP}

In this section, we provide an overview of our two phase approach for using VAE for AQP.
This requires solving a number of theoretical and practical challenges such as input encodings and approximation errors due to model bias.



\begin{figure}[h!]
	\centering
	\includegraphics[scale=0.3]{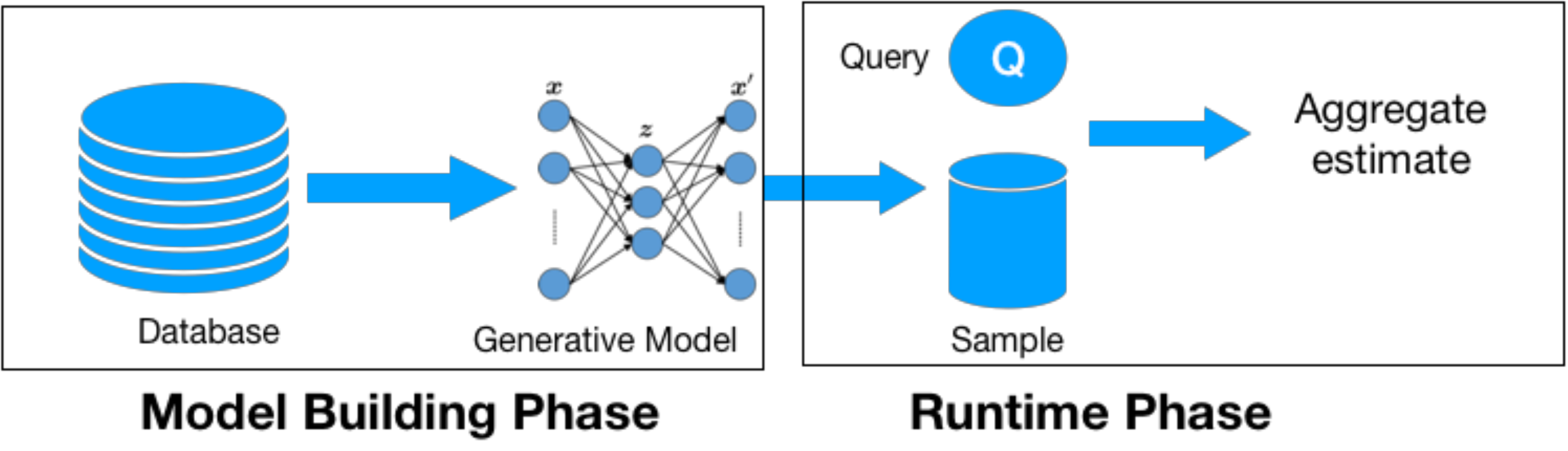}
	\caption{Two Phase Approach for DL based AQP}
	\label{fig:twoPhasedApproach}
\end{figure}

\stitle{Our Approach.}
Our proposed approach proceeds in two phases.
In the \emph{model building} phase, we train a deep generative model $M_{R}$ over the dataset $R$ such that it learns the underlying data distribution.
In this section, we assume that a single model is built for the entire dataset
that we relax in Section~\ref{sec:modelEnsembles}.
Once the DL model is trained, it can act as a succinct representation of the dataset.
In the \emph{run-time} phase, the AQP system uses the DL model to \emph{generate} samples $S$ from the underlying distribution.
The given query is rewritten to run on $S$.
The existing AQP techniques could be transparently used to generate the aggregate estimate.
Figure~\ref{fig:twoPhasedApproach} illustrates our approach.

\subsection{Using VAE for AQP}
\label{subsec:VAEForAQP}
In this subsection, we describe how to train a VAE over relational data and use it for AQP.

\stitle{Input Encoding.}
In contrast to homogeneous domains such as images and text, relations often consist of mixed data types that could be discrete or continuous.
The first step is to represent each tuple $t$ as a vector of dimension $d$.
For ease of exposition, we consider one-hot encoding and describe other effective encodings in Section~\ref{subsec:practicalTips}.
One-hot encoding represents each tuple as a $d = \sum_{i=1}^{m} |Dom(A_i)|$ dimensional vector
where the position corresponding to a given domain value is set to 1.
Each tuple in a relation $R$ with two binary attributes $A_1$ and $A_2$, is represented as a $4$ dimensional binary vector.
A tuple with $A_1=0,A_2=1$ is represented as $[1, 0, 0, 1]$ while a tuple with $A_1=1,A_2=1$ is represented as $[0, 1, 0, 1]$.
This approach is efficient for small attribute domains but becomes cumbersome if a relation has millions of distinct values.

\stitle{Model Building and Sampling from VAE.}
Once all the tuples are encoded appropriately, we could use VAE to learn the underlying distribution.
We denote the size of the input and latent dimension by $d$ and $d'$ respectively.
For one hot encoding, $d=\sum_{i=1}^{m} |Dom(A_i)|$.
As $d'$ increases, it results in more accurate learning of the distribution at the cost of a larger model.
Once the model is trained, it could be used to generate samples $X'$.
The randomly generated tuples often share similar statistical properties to tuples
sampled from the underlying relation $R$ and hence are a viable substitute for $R$.
One could apply the existing AQP mechanisms on the generated samples and use it to generate aggregate estimates along with confidence intervals.

The sample tuples are generated as follows:
we generate samples from the latent space $z$ and then apply the decoder network to convert points in latent space to tuples.
Recall from Section~\ref{sec:background} that the latent space is often a probability distribution that is easy to sample such as Gaussian.
It is possible to speed up the sampling from arbitrary Normal distributions using the reparameterization trick.
Instead of sampling from a distribution $N(\mu, \sigma)$, we could sample from the standard Normal distribution $N(0, 1)$ with zero mean and unit variance.
A sample $\epsilon$ from $N(0, 1)$ could be converted to a sample $N(\mu, \sigma)$ as $z = \mu + \sigma \odot \epsilon$.
Intuitively, this shifts $\epsilon$ by the mean $\mu$ and scales it based on the variance $\sigma$.

\subsection{Handling Approximation Errors.}
\label{subsec:approximationErrors}

We consider approximation error caused due to model bias
and propose an effective rejection sampling to mitigate it.

\stitle{Sampling Error.}
Aggregates estimated over the sample could differ from the exact results computed over the entire dataset
and their difference is called the sampling error.
Both the traditional AQP and our proposed approach suffer from sampling error.
The techniques used to mitigate it - such as increasing sample size - can also be applied to the samples from the generative model.

\stitle{Errors due to Model Bias.}
Another source of error is sampling bias.
This could occur when the samples are not representative of the underlying dataset and do not approximate its data distribution appropriately.
Aggregates generated over these samples are often biased and need to be corrected.
This problem is present even in traditional AQP~\cite{mozafari2015handbook} and mitigated through
techniques such as importance weighting~\cite{garofalakis2001approximate} and bootstrapping~\cite{efron1994introduction,mozafari2015handbook}.

Our proposed approach also suffers from sampling bias due to a subtle reason.
Generative models learn the data distribution which is a very challenging problem - especially in high dimensions.
A DL model learns an approximate distribution that is \emph{close enough}.
Uniform samples generated from the approximate distribution would be biased samples from the original distribution resulting in biased estimates.
%
%
As we shall show later in the experiments,
it is important to remove or reduce the impact of model bias to get accurate estimates.
Bootstrapping is not applicable as it often works by \emph{resampling} the sample data and performing inference on the sampling distribution from them.
Due to the biased nature of samples, this approach provides incorrect results~\cite{efron1994introduction}.
It is challenging to estimate the importance weight of a sample generated by VAE.
Popular approaches such as IWAE~\cite{burda2015importance} and AIS~\cite{neal2001annealed} do not provide strong bounds for the estimates.

\stitle{Rejection Sampling.}
We advocate for a rejection sampling based approach~\cite{grover2018variational,cochran2007sampling}
that has a number of appealing properties and is well suited for AQP.
Intuitively, rejection sampling works as follows.
Let $x$ be a sample generated from the VAE model with probabilities $p(x)$ and $q(x)$ from the original and approximate probability distributions respectively.
We accept the sample $x$ with probability $\frac{p(x)}{M \times q(x)}$ where $M$ is a constant upper bound on the ratio $p(x)/q(x)$ for all $x$.
We can see that the closer the ratio is to 1, the higher the likelihood that the sample is accepted.
On the other hand, if the two distributions are far enough, then a larger fraction of samples will be rejected.
One can generate arbitrary number of samples from the VAE model, apply rejection sampling on them
and use the accepted samples to generate unbiased and accurate aggregate estimates.

In order to accept/reject a sample $x$, we need the value of $p(x)$.
Estimating this value - such as by going to the underlying dataset - is very expensive and defeats the purpose of using generative models.
A better approach is to approximately estimate it purely from the VAE model.

\stitle{Variational Rejection Sampling Primer.}
We leverage an approach for variational rejection sampling that was recently proposed in~\cite{grover2018variational}.
For the sake of completeness, we describe the approach as applied to AQP.
Please refer to~\cite{grover2018variational} for further details.
Sample generation from VAE takes place in two steps.
First, we generate a sample $z$ in the latent space using the variational posterior $q(z|x)$ and
then we use the decoder to convert $z$ into a sample $x$ in the original space.
In order to generate samples from the true posterior $p(z|x)$, we need to accept/reject sample $z$ with acceptance probability
\begin{equation}
	a(z|x, M) = \frac{p(z|x)}{ M \times q(z|x)}
	\label{eq:acceptanceBasic}
\end{equation}
where $M$ is an upper bound on the ratio $p(z|x) / q(z|x)$.
Estimating the true posterior $p(z|x)$ requires access to the dataset and is very expensive.
However, we do know that the value of $p(x, z)$ from the VAE is within a constant normalization factor $p(x)$ as $p(z|x) = \frac{p(x, z)}{p(x)}$.
Thus, we can redefine Equation~\ref{eq:acceptanceBasic} as
\begin{equation}
	a(z|x, M') = \frac{p(x, z)}{ M \times p(x) \times q(z|x)} = \frac{p(x, z)}{M' \times q(z|x)}
	\label{eq:acceptanceAdvanced}
\end{equation}

We can now conduct rejection sampling if we know the value of $M'$.
First, we generate a sample $z$ from the variational posterior $q(z|x)$.
Next, we draw a random number $U$ in the interval $[0, 1]$ uniformly at random.
If this number is smaller than the acceptance probability $a(z|x, M')$, then we accept the sample and reject it otherwise.
That way the number of times that we have to repeat this process until we accept a sample is itself a random variable with geometric distribution p = $P(U \leq a(z|x,M'))$; $P(N=n)=(1-p)^{n-1}p, n \geq 1$. Thus on average the number of trials required to generate a sample is $E(N) = 1/p$. By a direct calculation it is easy to show \cite{cochran2007sampling} that $p=1/M'$.
We set the value of $M'$ as
	$M' = e^{-T}$
where $T \in [-\infty, +\infty]$ is an arbitrary threshold function.
This definition has a number of appealing properties.
First, this function is differentiable and can be easily plugged into the VAE's objective function
thereby allowing us to learn a suitable value of $T$ for the dataset during training~\cite{grover2018variational}.
Please refer to Section~\ref{sec:experiments} for a heuristic method for setting appropriate values of $T$ during model building and sample generation.
Second, the parameter $T$ when set, establishes a trade-off between computational efficiency and accuracy.
If $T \rightarrow +\infty$,  then every sample is accepted (\ie no rejection) resulting into fast sample generation at the expense of the quality of the approximation to the true underlying distribution.
In contrast when $T \rightarrow -\infty$, we ensure that almost every sample is guaranteed to be from the true posterior distribution, by making the acceptance probability small and as a result increasing sample generation time.
Since $a$ should be a probability we change equation Equation~\ref{eq:acceptanceAdvanced} to:
\begin{equation}
	a(z|x, M') = \min \Big[ 1, \frac{p(x, z)}{M' \times q(z|x)} \Big]
	\label{eq:acceptanceFinal}
\end{equation}

\subsection{Towards Accuracy Guarantees}
\label{subsec:towardsAccuracyGuarantees}

\add{As mentioned in Section~\ref{sec:introduction},
our approach is complementary to traditional AQP system.
Our objective is to design a lightweight deep generative model that could be used
to obtain quick-and-dirty aggregate estimates that are often sufficient for preliminary data exploration.
Once the user has identified promising queries that requires provable guarantees,
we can defer traditional AQP techniques or even obtain exact answers.
In this subsection, we describe an initial approach for obtaining the accuracy guarantees.
We would like to note that developing a framework to quantify approximation errors of AQP based on
deep generative models is a challenging problem and a focus of our future research.}

\add{\stitle{Eliminating Model Bias.}
Recall from Section~\ref{subsec:approximationErrors} that
approximation errors incurred in our approach are due to model bias and sampling error.
If the model bias is eliminated, then our problem boils down to the traditional AQP setting.
We could readily leverage
the rich set of accuracy guarantees and confidence intervals developed for handling sampling error.
This is achieved by setting $T = -\infty$ and applying variational rejection sampling (VRS).
However, this comes with a large computational cost
whereby the vast majority of generated tuples are rejected.
Ideally, we would like a granular accuracy-computation tradeoff.
Increasing $T$ improves the sampling efficiency at the cost of model bias.}

\add{\stitle{Distribution Testing.}
We adapt techniques designed for high-dimensional two-sample hypothesis testing~\cite{rosenbaum2005exact,modelTwoSampleTests} for choosing appropriate $T$.
Suppose we are given two sets of uniform random samples $S_{D}$ and $S_{M}$
from the original dataset and the learned model respectively.
Let $|S_{D}| = |S_{M}|$.
Suppose that these samples were drawn from probability distributions $P_{D}$ and $P_{M}$.
If we can ascertain that the two distributions are the same (i.e. $P_{D} = P_{M}$),
by using the corresponding samples, then we can safely ignore the issue of model bias.
This is achieved by testing the null hypothesis $H_0: P_{D} = P_{M}$.}

\add{There are two factors that makes this challenging:
high-dimensionality and test-statistics for AQP.
First, we train VAE model by transforming tuples into a vector
whose dimensionality ranges in the thousands.
Classical tests such as Kolmogrov-Smirnov
are not suitable for testing such high dimensional distributions.
Second, hypothesis testing methods rely on a test statistic that is a function of $S_D$ and $S_M$
that could be used to distinguish $P_{D}$ and $P_{M}$.
For example, a simple test statistic is to choose an aggregate query such as
estimating the average value of some attribute $A_i$.
If the average of $A_i$ computed over $S_D$ and $S_M$ deviates beyond certain threshold
we can reject the null hypothesis.
However, this is not appropriate for our scenario.
We wish to test the null hypothesis for arbitrary aggregate queries.
The way out of this conundrum is to use \emph{Cross-Match Test}~\cite{rosenbaum2005exact,modelTwoSampleTests}.
}

\add{\stitle{Cross-Match Test for AQP.}
We begin by projecting tuples in $S_D$ and $S_M$ into the latent space of VAE using the encoder.
We abuse the notation by representing the projected tuples as $S_D$ and $S_M$.
Let $Z = S_D \cup S_M$.
We associate a label of $0$ if tuple $t \in S_D$ and a label of $1$ if $t \in S_M$.
We construct a \emph{complete} graph where each node corresponds to a tuple in $Z$
while the edge corresponds the Euclidean distance between
the latent space representation of the corresponding tuples.
We then compute a minimum weight perfect matching using the Blossom algorithm~\cite{edmonds1965paths}.
The output is a collection of non-overlapping pairs of tuples.
Consider a specific pair of tuples $(Z_i, Z_j)$.
There are three possibilities: both tuples are from $S_D$, both tuples are from $S_M$
or one each from $S_D$ and $S_M$.
Let $a_{D,D}, a_{M,M}, a_{D,M}$ be the frequency of pairs from the matching of these three categories.
The cross-match test~\cite{rosenbaum2005exact,modelTwoSampleTests} specifies $a_{D,M}$ as the test statistic.
Let $\eta=a_{D,D} + a_{M,M} + a_{D,M}$.
We accept or reject the null hypothesis based on the probability computed as }
\add{
\begin{equation}
	\frac{2^{a_{D,M}} \times \eta!}{ {\eta \choose |S_D|} (a_{D,D})! (a_{M,M})! (a_{M,D})!}
	\label{eq:crossMatchTestStatistic}
\end{equation}
}

\subsection{Variational Autoencoder AQP Workflow}
\label{subsec:puttingItAllTogether}
Algorithm~\ref{alg:VAEforAQP} provides the pseudocode for the overall workflow of performing AQP using VAE.
In the model building phase, we encode the input relation $R$ using an appropriate mechanism (see Section \ref{subsec:practicalTips}).
The VAE model is trained on the encoded input and stored along with appropriate metadata.
\add{
During the runtime phase, we generate sample $S_M$ from VAE
using variational rejection sampling with $T=0$.
We then apply the hypothesis testing to ensure that the two distributions cannot be distinguished.
If the null hypothesis is rejected, we generate a new sample $S_D$ with a lower value of $T$.
This will ensure that the model bias issue is eliminated.
One can then apply existing techniques for generating approximation guarantees and confidence intervals.
Note that we use the VAE model for data exploration only after it passed the hypothesis testing.
}

\begin{algorithm}[h!]
	\caption{AQP using VAE}
	\label{alg:VAEforAQP}
	\begin{algorithmic}[1]
		\STATE \textbf{Input:} VAE model $\mathcal{V}$
		\STATE $T=0$, $S_{D}$ = sample from $D$,
		\STATE $ S_z = \{\}$ \qquad {//set of samples}
		\WHILE{samples are still needed}
			\STATE Sample $z \sim q(z|x)$
			\STATE Accept or reject $z$ based on Equation~\ref{eq:acceptanceFinal}
			\STATE \textbf{If} $z$ is accepted, $S_z = S_z \cup \{z\}$
		\ENDWHILE
		\STATE $S_M=$Decoder($S_z)$ {// Convert samples to original space}
		\STATE Test null hypothesis $H_0: P_S = P_D$ using Equation~\ref{eq:crossMatchTestStatistic}
		\IF {$H_0$ is rejected}
			\STATE $T = T - 1$
			\STATE Goto Step 3
		\ENDIF
		\STATE \textbf{Output:} Model $\mathcal{V}$ and $T$
	\end{algorithmic}
\end{algorithm}

\subsection{Making VAE practical for relational AQP}
\label{subsec:practicalTips}
In this subsection, we propose two practical improvements for training VAE for AQP over relational data.

\stitle{Effective Input Encoding.}
One-hot encoding of tuples is an effective approach for relatively small attribute domains.
If the relation has millions of distinct values, then it causes two major issues.
First, the encoded vector becomes very sparse resulting in poor performance~\cite{krishnan2016inference}.
Second, it increases the number of parameters learned by the model thereby increasing the model size and the training time.


A promising approach to improve one-hot encoding is to make the representation denser using \emph{binary encoding}.
Without loss of generality, let the domain $Dom(A_j)$ be its zero-indexed position $[0, 1, \ldots, |Dom(A_j)|-1]$.
We can now concisely represent these values using $\lceil \log_2 |Dom(A_j)| \rceil$ dimensional vector.
Once again consider the example $Dom(A_j) = \{Low, Medium, High\}$.
Instead of representing $A_j$ as a 3-dimensional vectors (\ie $001,010,100$),
we can now represent them in $\lceil \log_2(3) \rceil = 2$-dimensional vector \ie $\eta(Low)=00, \eta(Medium)=01, \eta(High)=10$.
This approach is then repeated for each attribute resulting a $d = \sum_{i=1}^{n} \lceil \log_2 |Dom(A_i)| \rceil$-dimensional vector (for $n$ attributes)
that is exponentially smaller and denser than the one-hot encoding that requires $\sum_{i=1}^{n} |Dom(A_i)|$ dimensions.

\stitle{Effective Decoding of Samples.}
Typically, samples are obtained from VAE in two steps:
(a) generate a sample $z$ in the latent space \ie $z \sim q(z|x)$ and
(b) generate a sample $x'$ in the original space by passing $z$ to the decoder.
While this approach is widely used in many domains such as images and music, it is not appropriate for databases.
Typically, the output of the decoder is stochastic.
In other words, for the same value of $z$, it is possible to generate multiple reconstructed tuples from the distribution $p(x|z)$.
However, blindly generating a random tuple from the decoder output could return an invalid tuple.
For images and music, obtaining incorrect values for a few pixels/notes is often imperceptible.
However, getting an attribute wrong could result in a (slightly) incorrect estimate
Typically, the samples generated are often more correct than wrong.
We could minimize the likelihood of an aberration by generating multiple samples for the same value of $z$.
In other words, for the same latent space sample $z$, we generate multiple samples $X'=\{x'_1, x'_2, \ldots, \}$ in the tuple space.
These samples could then be aggregated to obtain a single sample tuple $x'$.
The aggregation could be based on max (\ie for each attribute $A_j$, pick the value that occurred most in $X'$)
or weighted random sampling (\ie for each attribute $A_j$, pick the value based on the frequency distribution of $A_j$ in $X'$).
Both these approaches provide sample tuples that are much more robust resulting in better accuracy estimates.

%% file: modelEnsembles.tex
\section{AQP using Multiple VAEs}
\label{sec:modelEnsembles}

So far we have assumed that a single VAE model is used to learn the data distribution.
As our experimental results show, even a single model could generate effective samples for AQP.
However, it is possible to improve this performance and generate better samples.
One way to accomplish this is to split the dataset into say $K$ non-overlapping partitions and learn a VAE model for each of the partitions.
Intuitively, we would expect each of the models to learn the finer characteristics of the data from the corresponding partition
and thereby generate better samples for that partition.
In this section, we investigate the problem of identifying the optimal set of $K$ partitions for building VAE models.

\subsection{Problem Setup}
\label{subsec:ensembleProblemSetup}
Typically, especially in OLAP settings, tuples are grouped according to hierarchies on given attributes.
Such hierarchies reflect meaningful groupings which are application specific such as for example location, product semantics, year, etc.
Often, these groupings have a semantic interpretation and building models for such groupings makes more sense than doing so on an arbitrary subset of the tuples in the dataset.
As an example, the dataset could be partitioned based on the attribute Country
such that all tuples belonging to a particular country is an atomic group.
We wish to identify $K$ non-overlapping groups of countries such that a VAE model is trained on each group.

More formally, let $G = \{g_1, g_2, \ldots, g_l\}$ be the set of existing groups with $g_i \subseteq R$
such that $\cup_{i=1}^{l} g_i = R$.
We wish to identify a partition $S = \{s_1, \ldots, s_K\}$ of $R$
where $s_i \subseteq G$ and $s_i \cap s_j = \emptyset$ when $i \neq j$.
Our objective is to group these $l$ subsets into $K$ non-overlapping partitions such that
the aggregate error of the VAEs over these partitions is minimized.

Efficiently solving this problem involves two steps:
(a) given a partition, a mechanism to {\em estimate} the error of $K$ VAEs trained over the partition \emph{without} conducting the actual training and
(b) an algorithm that uses (a) to identify the best partition over the space of partitions.
Both of these challenges are non-trivial.

\subsection{Bounding VAE Errors}
\label{subsec:boundingVAEErrors}

\stitle{Quantifying VAE Approximation.}
The parameters of VAE are learned by optimizing an evidence lower bound (ELBO) given by
$$E[\log P(X|z)]-D_{KL}[Q(z|X)||P(z)]$$
(from Equation~\ref{eq:var2})
which is a tight bound on the marginal log likelihood.
ELBO provides a meaningful way to measure the distribution approximation by the VAE.
Recall from Section~\ref{subsec:approximationErrors} that we perform rejection sampling on the VAE
that results in a related measure we call R-ELBO (resampled ELBO) defined as
$$E[\log P(X|z)]-D_{KL}[R(z|X,T)||P(z)]$$
where $R(z|X,T)$ is the resampled distribution for a user-specified threshold of $T$.
Given two VAEs trained on the same dataset for a fixed value of $T$, the VAE with lower R-ELBO provides a better approximation.

\stitle{Bounding R-ELBO for a Partition.}
Let us assume that we will train a VAE model for each of the atomic groups $g_j \in G$.
We train the model using variational rejection sampling~\cite{grover2018variational} for a fixed $T$ and compute its R-ELBO.
In order to find the optimal partition, we have to compute the value of R-ELBO for arbitrary subsets $s_i = \{g_{i1}, \ldots,\} \subseteq G$.
The naive approach would be to train a VAE on the union of the data from atomic groups in $s_i$ which is time consuming.
Instead, we empirically show that it is possible to bound the R-ELBO of VAE trained on $s_i$
if we know the value of R-ELBO of each of $g_{i1}, \ldots, $.
Let $f(\cdot)$ be such a function.
In this paper, we take a conservative approach and bound it by sum $f(r_1, r_2, \ldots, r_k)  = \sum_{i=1}^{k} r_i$ where $r_i$ is the R-ELBO for group $g_i$.
In other words, $f(\cdot)$ bounds the R-ELBO of VAE trained $\cup_{i=1}^{k} g_i$ by $\sum_{i=1}^{k} r_i$.
It is possible to use other functions that provide tighter bounds.

\stitle{Empirical Validation.}
We empirically validated the function $f(\cdot)$ on a number of datasets under a variety of settings.
Table~\ref{tbl:elboValidation} show the results for Census and Flights dataset that has been widely used in prior work on AQP such as~\cite{kulessa2018model,Galakatos:2017:RRA:3115404.3115418,getoor2001selectivity}.
Please refer to Section~\ref{sec:experiments} for a description of the two datasets.
We obtained similar results for other benchmark datasets.
For each of the datasets, we constructed multiple atomic groups for different categorical attributes.
For example, one could group the Census dataset using attributes such as gender, income, race etc.
We ensured that each of the groups are at least 5\% of the data set size to avoid outlier groups and if necessary merged smaller groups into a miscellaneous group.
We trained a VAE model on each of the groups for different values of $T$ using variational rejection sampling and computed their R-ELBO.
We then construct all pairs, triples, and other larger subsets of the groups
and compare the bound obtained by $f(\cdot)$ with the actual R-ELBO value of the VAE trained on the data of these subsets.
For each dataset, we evaluated 1000 randomly selected subsets and report the fraction in which the bound was true.
As is evident in table \ref{tbl:elboValidation} the bound almost always holds.

{\tiny
\begin{table}[h]
\begin{center}
\caption{Empirical validation of R-ELBO Bounding}
\label{tbl:elboValidation}
\begin{tabular}{|l|l|l|l|}
	\hline
	\textbf{Dataset} & $T=-10$ & $T=0$ & $T=+10$ \\ \hline
	Census  &  0.992      &  0.997   &   0.996     \\ \hline
	Flights &  0.961      &  0.972   &   0.977     \\ \hline
\end{tabular}
\end{center}
\end{table}
}

\subsection{Choosing Optimal Partition}
\label{subsec:choosingOptimalPartition}

In this section we assume we are provided with the value of R-ELBO for each of the groups $g_i \in G, 1 \leq i \leq l$, a bounding function $f(\cdot)$ and a user specified value $K$.
We propose an algorithm that optimally splits a relation $D$ into $K$ non overlapping partitions $S = \{s_1, \ldots, s_K\}$
where $s_i \subseteq G$ and $s_i \cap s_j = \emptyset$ when $i \neq j$.
The key objective is to choose the split $S$ in such a way that the $\sum_{i=1}^{K} \text{R-ELBO}(s_i)$ is minimized.
Note that there are $K^l$ possible partitions and exhaustively enumerating and choosing the best partition is often infeasible.
R-ELBO($g$) corresponds to the actual R-ELBO for atomic groups $g \in G$
while for $s_i \subseteq G$, this is estimated using the bounding function $f(s_i)$.
We investigate scenarios that occur in practice.

\stitle{Optimal Partition using OLAP Hierarchy.}
In OLAP settings, tuples are grouped according to hierarchies on given attributes that reflect meaningful semantics.
We assume the availability of an OLAP hierarchy in the form of a tree
where the leaf node corresponds to the atomic groups (\eg Nikon Digital Cameras)
while the intermediate groups correspond to product semantics (\eg Digital Camera $\rightarrow$ Camera $\rightarrow$ Electronics and so on).
We wish to build VAE on meaningful groups of tuples by constraining $s_i$ to be selected from the leafs or intermediate nodes,
be mutually exclusive and have the least aggregate R-ELBO score.
We observe that the selected nodes forms a tree cut that partitions the OLAP hierarchy into $K$ disjoint sub-trees.

Let us begin by considering the simple scenario where the OLAP hierarchy is a binary tree.
Let $h$ denote an arbitrary node in the hierarchy with \emph{left(h)} and \emph{right(h)}
returning the left and right children of $h$ if they exist.
We propose a dynamic programming algorithm to compute the optimal partition.
We use the table $Err[h, k]$ to denote aggregate R-ELBO of splitting the sub-tree rooted at node $h$ using at most $k$ partitions where $k \leq K$.
The base case $k=1$ is simply building the VAE on all the tuples falling under node $h$.
When $k > 1$, we evaluate the various ways to split $h$ such that the aggregate R-ELBO is minimized.
For example, when $k=2$, there are two possibilities.
We could either not split $h$ or build two VAE models over \emph{left(h)} and \emph{right(h)}.
The optimal decision could be decided by choosing the option with least aggregate error.
In general, we consider all possible ways of apportioning $K$ between the left and right sub-trees of $h$
and pick the allocation resulting in least error.
The recurrence relation is specified by,
\begin{equation}
	\label{eq:DPRecurrenceBinary}
	Err[h, k] =
		\begin{cases}
			\mbox{R-ELBO(h)}  & \mbox{ if } $k=1$\\
			\min_{1 \leq i \leq k} ( Err[\text{left(h)}, i] + & \\
			Err[\text{right(h)}, k - i] ) & \mbox{ otherwise}
		\end{cases}
\end{equation}

The extension to non-binary trees is also straightforward.
Let $C=\{c_1, \ldots, c_j\}$ be the children of node $h$.
We systematically partition the space of children into various groups of two and identify the best partitioning that gives the least error (eq. \ref{eq:DPRecurrenceGeneral}).
A similar dynamic programming approach was also used for constructing histograms over hierarchical data
in~\cite{reiss2006compact}.
\begin{equation}
	\label{eq:DPRecurrenceGeneral}
	Err[h, k] =
		\begin{cases}
			\mbox{R-ELBO(h)}  & \mbox{ if } $k=1$\\
			\min_{1 \leq i \leq k} ( Err[\{c_1, \ldots, c_{j/2}\}, i] + & \\
			Err[\{c_{j/2+1}, \ldots, c_j\}, k - i] ) & \mbox{ otherwise}
		\end{cases}
\end{equation}

\stitle{Scenario 2: Partitioning with Contiguous Atomic Groups.}
Given the atomic groups $G = \{g_1, \ldots, g_l\}$, a common scenario is to partition them into $K$ contiguous subsets.
This could be specified as $K+1$ integers $1=b_1 \leq b_2 \ldots \leq b_{K+1}=l$
where the boundary of the $i$-th subset is specified by $[b_{i}, b_{i+1}]$ and consists of a set of atomic groups $\{g_{b_i}, \ldots,  g_{b_{i+1}}]$.
This is often desirable when the underlying attribute has a natural ordering such as year.
So we would prefer to train VAE models over data from consecutive years
such as $\{2016-2017, 2018-2019\}$ instead of arbitrary groupings such as $\{ \{2016, 2018\}, \{2017, 2019\}\}$.
This problem could be solved in near linear time (\ie $O(l)$) by using the approach first proposed in~\cite{Guha:2002:FAH:543613.543637}.
The key insight is the notion of \emph{sparse interval set system} that could be used to express any interval using a bounded number of sparse intervals.
The authors then use a dynamic programming approach on the set of sparse intervals to identify the best partitioning.


\add{In practice, $K$ is often determined by various other factors such as space budget for
persisting the generative models. Identifying $K$ automatically is an interesting orthogonal problem.
Our bounding function for R-ELBO has a natural monotonic property.
We empirically found that common heuristics for selecting
number of clusters such as Elbow method~\cite{han2011data} works well for our purpose.
}

%% file: experiments.tex
\section{Experiments}
\label{sec:experiments}

We conduct a comprehensive set of experiments
and demonstrate that VAE (and deep generative models) are a promising mechanism for AQP.
We reiterate that our proposed approach is an alternate way for generating samples, albeit very fast.
Most of the prior work for improving AQP estimates could be transparently used on the samples from VAE.

\begin{figure*}[!ht]
     \begin{minipage}[t]{0.32\linewidth}
         \centering
         \includegraphics[width =\textwidth]{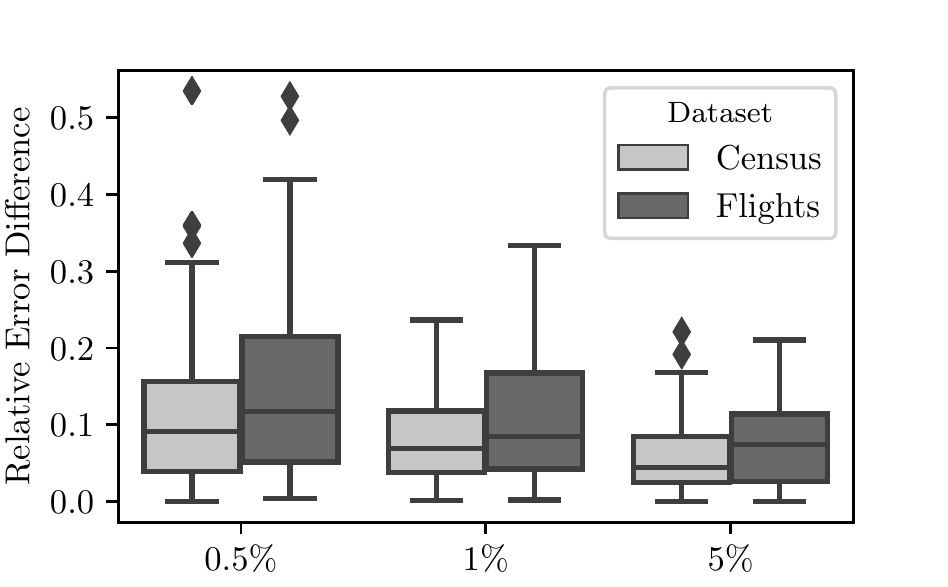}
         \caption{Varying Sample Size}
         \label{fig:VAEModelQuality}
     \end{minipage}
     \hspace{1mm}
     \begin{minipage}[t]{0.32\linewidth}
         \centering
         \includegraphics[width =\textwidth]{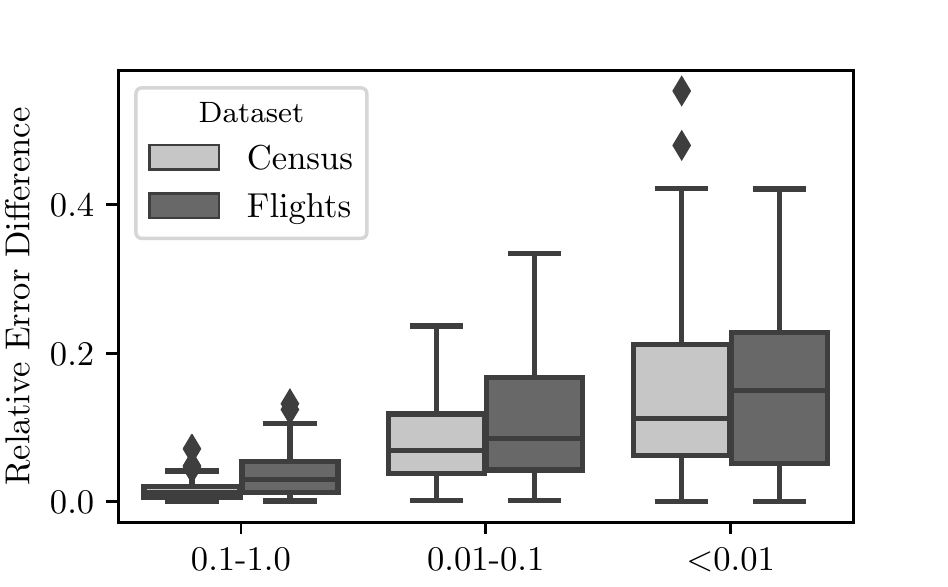}
         \caption{Varying Query Selectivity}
         \label{fig:VAESelectivity}
     \end{minipage}
     \hspace{1mm}
     \begin{minipage}[t]{0.32\linewidth}
         \centering
         \includegraphics[width =\textwidth]{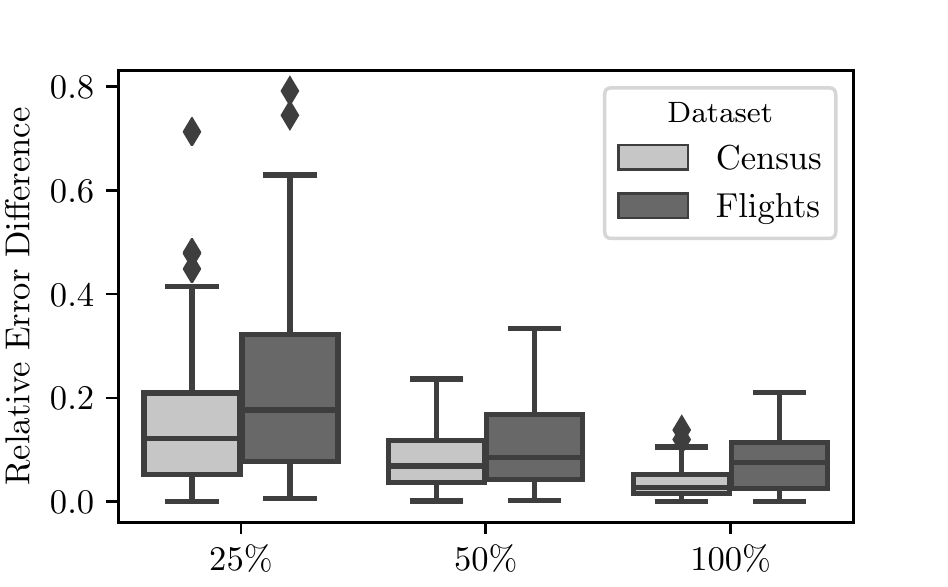}
         \caption{Varying Latent Dimension}
         \label{fig:VAELatentDimension}
     \end{minipage}
 \end{figure*}

 \begin{figure*}[!ht]
     \begin{minipage}[t]{0.32\linewidth}
         \centering
         \includegraphics[width =\textwidth]{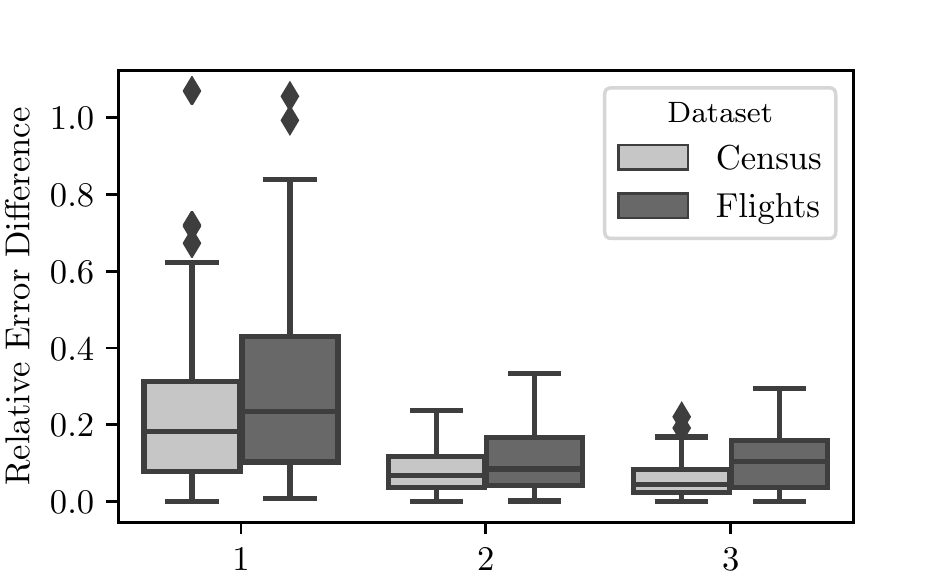}
         \caption{Varying Model Depth}
         \label{fig:VAEDepth}
     \end{minipage}
     \hspace{1mm}
     \begin{minipage}[t]{0.32\linewidth}
         \centering
         \includegraphics[width =\textwidth]{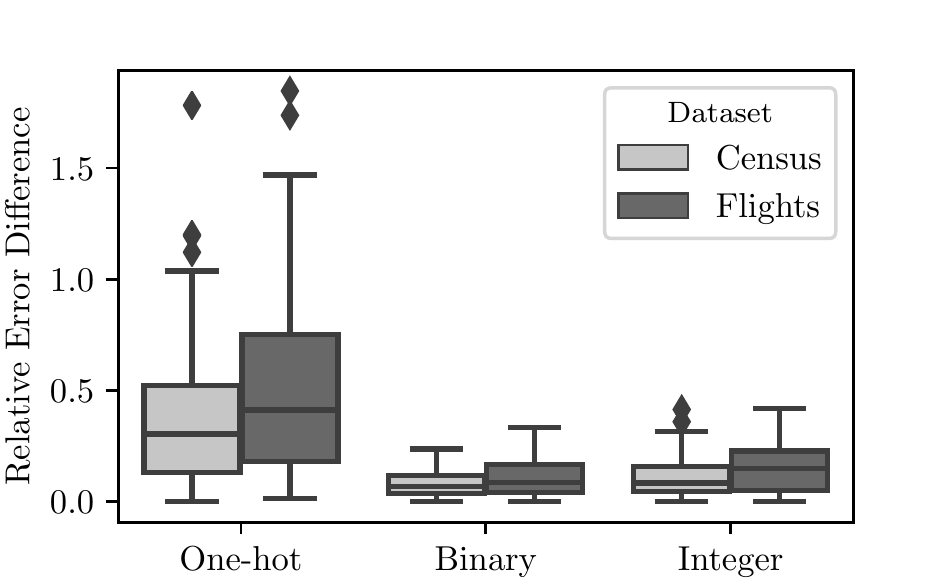}
         \caption{Varying Input Encoding}
         \label{fig:VAEInputEncoding}
     \end{minipage}
     \hspace{1mm}
     \begin{minipage}[t]{0.32\linewidth}
         \centering
         \includegraphics[width =\textwidth]{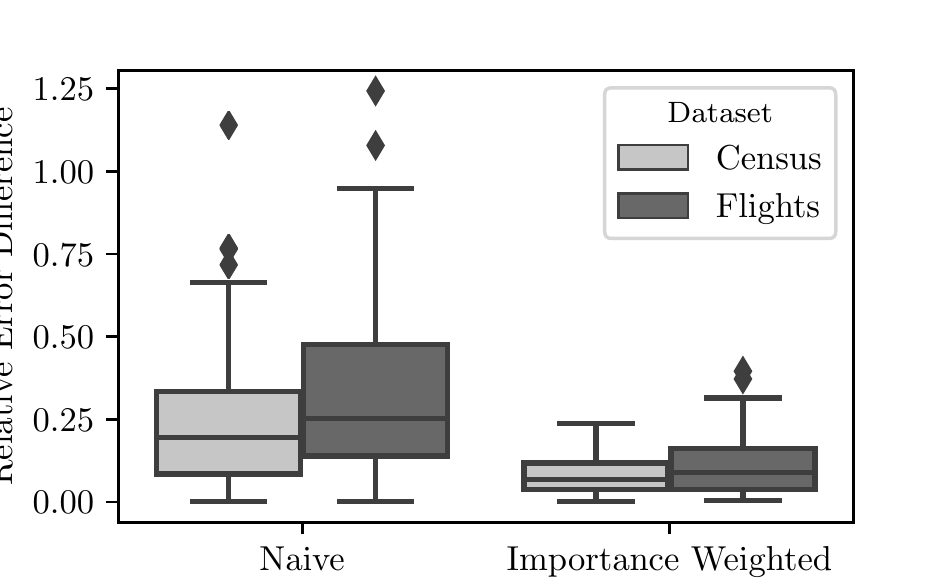}
         \caption{Varying Output Encoding}
         \label{fig:VAEOutputEncoding}
     \end{minipage}
 \end{figure*}

\subsection{Experimental Setup}
\label{subsec:expSetup}

\stitle{Hardware and Platform.}
All our experiments were performed on a server with 16 cores, 128 GB of RAM and NVidia Tesla K80 GPU.
We used PyTorch~\cite{paszke2017automatic} for training VAE and GAN, bnlearn~\cite{bnlearn} for learning Bayesian Networks and MSPN~\cite{molina2018mixed} for mixed sum-product networks (MSPN).

\stitle{Datasets.}
We conducted our experiments on two real-world datasets: Census~\cite{census} and Flights~\cite{flights,eichmann2018idebench}.
Both datasets have  complex correlated attributes and conditional dependencies that make AQP challenging.
The Census dataset has 8 categorical attributes and 6 numerical attributes and contains demographic and employment information.
The Flights dataset has 6 categorical and 6 numerical attributes and contains information about on-arrival statistics for the last few years.
We used the data generator from~\cite{eichmann2018idebench} to scale the datasets to arbitrary sizes while also ensuring that the relationships between attributes are maintained.
By default, our experiments were run on datasets with 1 million tuples.

\stitle{Deep Generative Models for AQP.}
In our experiments, we primarily focus on VAE for AQP as it is easy and efficient to train and generates realistic samples \cite{doersch2016tutorial}.
By default, our VAE model consists of a 2 layer encoder and decoder that are parameterized by Normal and Bernoulli distributions respectively.
We used binary encoding (Section~\ref{subsec:practicalTips}) for converting tuples into a representation consumed by the encoder.

In order to generate high quality samples, we use rejection sampling during both VAE training and sample generation albeit at different granularities.
During training, the value of threshold $T(x)$ is set for each tuple $x$
so that the acceptance probability of samples generated from $q(z|x)$ is roughly $0.9$ for most tuples.
We use the procedure from~\cite{grover2018variational}
 to generate a Monte Carlo estimate for $T(x)$ satisfying acceptance probability constraints.
While the trained model already produces realistic samples, we further ensure this by
performing rejection sampling with a fixed threshold $T$ (for the entire dataset) during sample generation (as detailed in Section~\ref{subsec:approximationErrors}).
There are many ways for choosing the value of $T$.
It could be provided by the user or chosen by cross validation such that it provides the best performance on query workload.
By default, we compute the value of $T$ from the final epoch of training as follows.
For each tuple $x$, we have the Monte-Carlo estimate $T(x)$.
We select the 90-th percentile of the distribution $T(x)$.
Intuitively, this ensures that samples generated for 90\% of the tuples would have acceptance probability of 0.9.
Of course, it is possible to specify different values of $T$ for queries with stringent accuracy requirements.
We used Wasserstein GAN as the architecture for generative adversarial networks \cite{Goodfellow-et-al-2016}.
We used entropy based discretization~\cite{dougherty1995supervised} for continuous attributes when training discrete Bayesian networks.
We used the default settings from~\cite{molina2018mixed} for training MSPN.

\stitle{Query Workload.}
We used IDEBench~\cite{eichmann2018idebench} to generate aggregate queries involving filter and group-by conditions.
We then selected a set of 1000 queries that are diverse in various facets such as
number of predicates, selectivity, number of groups, attribute correlation etc.

\stitle{Performance Measures.}
As detailed in Section~\ref{subsec:approximationErrors}, AQP using VAE introduces two sources of errors: sampling error and errors due to model bias.
The accuracy of an estimate could be evaluated by relative error (see Equation~\ref{eq:relErr}).
For each query in the workload, we compute the relative error over a fixed size sample (1\% by default)
obtained from the underlying dataset $R$ and the learned VAE model.
For a given query, the relative error difference (RED) computed as the absolute difference between the two relative errors
provides a meaningful way to compare them.
Intuitively, RED will be close to 0 for a well trained VAE model.
We repeat this process over 10 different samples and report the average results.
Given that our query workload has 1000 queries, we use box plots to concisely visualize the distribution of the relative error difference.
The middle line corresponds to the median value of the difference while the box boundaries correspond to the 25th and 75th percentiles.
The top and bottom whiskers are set to show the 95th and 5th percentiles respectively.

\subsection{Experimental Results}
\label{subsec:expResults}

\stitle{Evaluating Model Quality.}
In our first experiment, we demonstrate that VAE could meaningfully learn the data distribution and generate realistic samples.
Figure~\ref{fig:VAEModelQuality} shows the distribution of relative error differences for both datasets over the entire query workload for various sample sizes.
We can see that the differences are less than 1\% for almost all the cases for the Census dataset.
The flights dataset has many attributes with large domain cardinalities which makes learning the data distribution very challenging.
Nevertheless, our proposed approach is still within 3\% of the relative error obtained from the samples of $R$.

\stitle{Impact of Selectivity.}
In this experiment, we group the queries based on their selectivity and compute the relative error difference for each group.
As shown in Figure~\ref{fig:VAESelectivity}, the difference is vanishingly small for queries with large selectivities and slowly increases for decreasing selectivities.
In general, generating estimates for low selectivity queries is challenging for any sampling based AQP.
The capacity/model size constraints imposed on the VAE model could result in generating bad estimates for some queries with very low selectivities.
However, this issue could be readily ameliorated by building multiple VAE models that learn the finer characteristics of data minimizing such errors in these cases.

\stitle{Impact of Model Capacity and Depth.}
Figures~\ref{fig:VAELatentDimension} and~\ref{fig:VAEDepth} shows the impact of two important hyper parameters -
the number of latent dimensions and depth of the encoder and decoder.
We vary the latent dimension from 10\% to 100\% of the input dimension.
Large latent dimension results in an expressive model that can learn complex data distributions at the cost of increased model size and training time.
Increasing the depth results in a more accurate model but with larger model size and slower training time.
Empirically, we found that setting latent dimension size to 50\% (for binary encoding) and encoder/decoder network depth of 2 provides good results.

\stitle{Effectiveness of Input Encoding and Output Decoding.}
It is our observation that the traditional approach of one-hot encoding coupled with generating a single sample tuple for each sample from the latent space does not provide realistic tuples. It may be suitable for image data but certainly not suitable for relational data.
Figure~\ref{fig:VAEInputEncoding} shows how different encodings affect the generated samples.
For datasets such as Census where almost all attributes have small domain cardinality, all the three approaches provide similar results.
However, for the flights dataset where some attributes have domain cardinality in tens of thousands,
naive approaches such as one-hot encoding provides sub-optimal results.
This is due to the fact that there are simply too many parameters to be learnt and even a large dataset of 1 Million tuples is insufficient.
Similarly, Figure~\ref{fig:VAEInputEncoding} shows that our proposed decoding approach dramatically decreases the relative error difference making the approach suitable for relational data.
This is due to the fact that the naive decoding could produce unrealistic tuples that could violate common integrity constraints an effect
that is minimized when using our proposed decoding.

\begin{figure*}[!ht]
     \begin{minipage}[t]{0.32\linewidth}
         \centering
         \includegraphics[width =\textwidth]{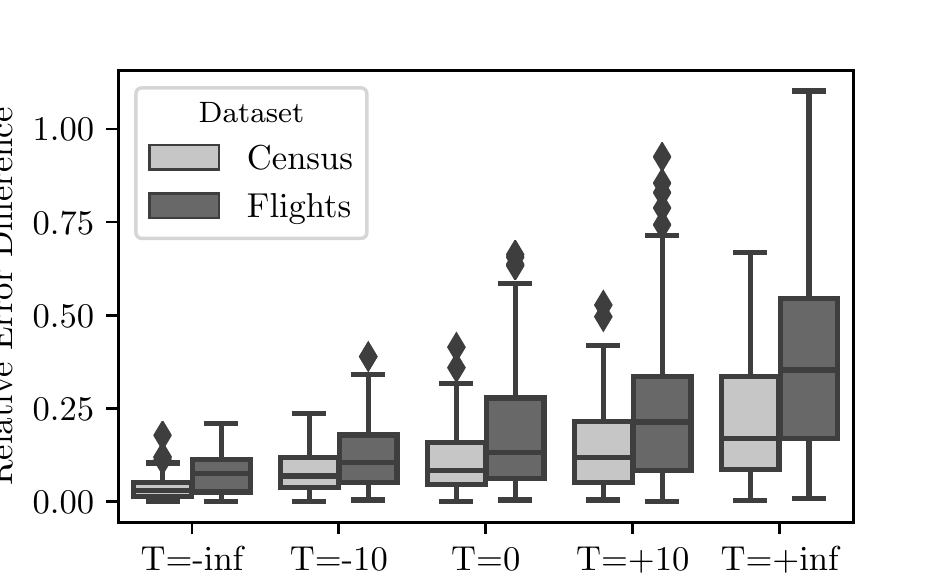}
         \caption{Varying $T$}
         \label{fig:VAERejectionSampling}
     \end{minipage}
     \hspace{1mm}
     \begin{minipage}[t]{0.32\linewidth}
         \centering
         \includegraphics[width =\textwidth]{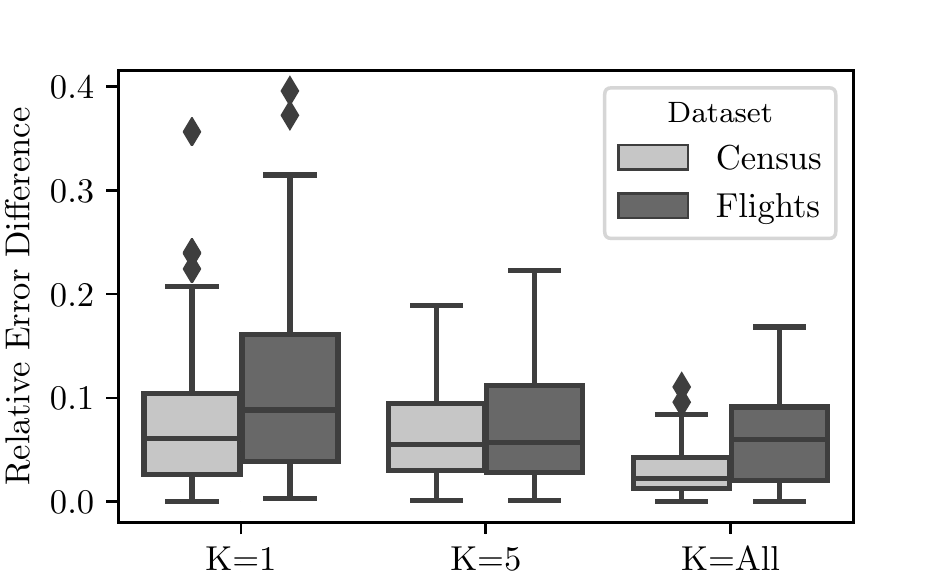}
         \caption{Varying $K$}
         \label{fig:oneVsManyVAE}
     \end{minipage}
     \hspace{1mm}
     \begin{minipage}[t]{0.32\linewidth}
         \centering
         \includegraphics[width =\textwidth]{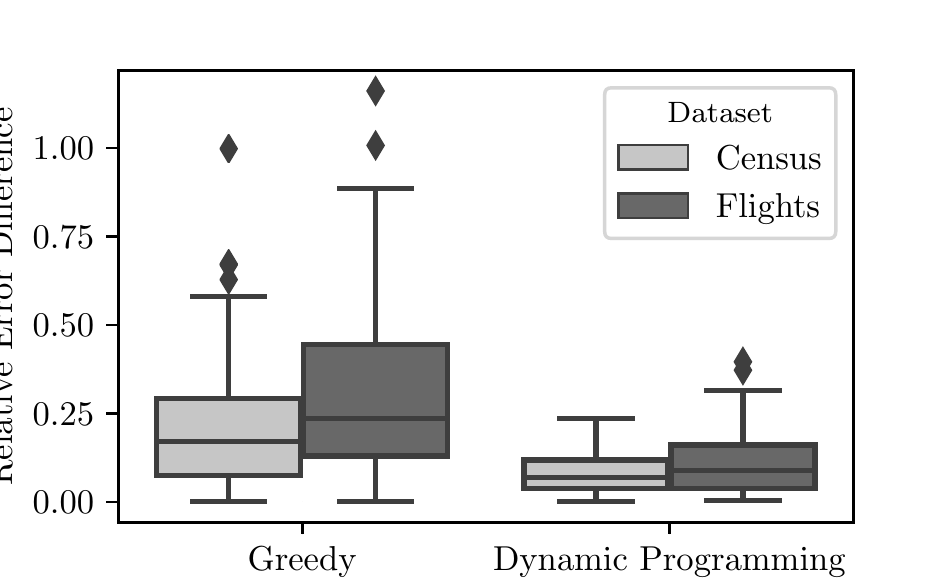}
         \caption{Partition Algorithms}
         \label{fig:oneVsManyVAEDPAlgo}
     \end{minipage}
 \end{figure*}

 \begin{figure*}[!ht]
     \begin{minipage}[t]{0.32\linewidth}
         \centering
         \includegraphics[width =\textwidth]{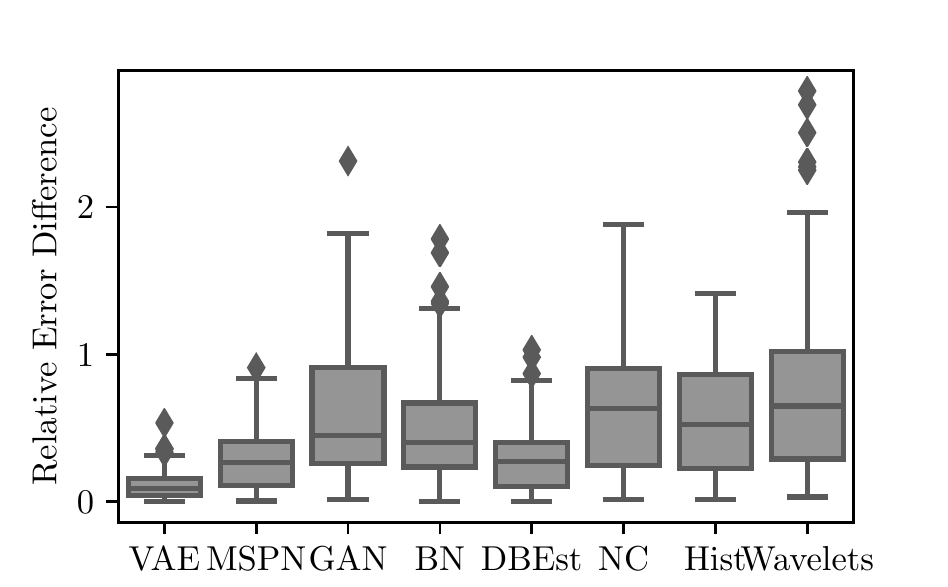}
         \caption{Performance of DL models for AQP}
         \label{fig:allDLModels}
     \end{minipage}
     \hspace{1mm}
     \begin{minipage}[t]{0.34\linewidth}
         \centering
         \includegraphics[width =\textwidth]{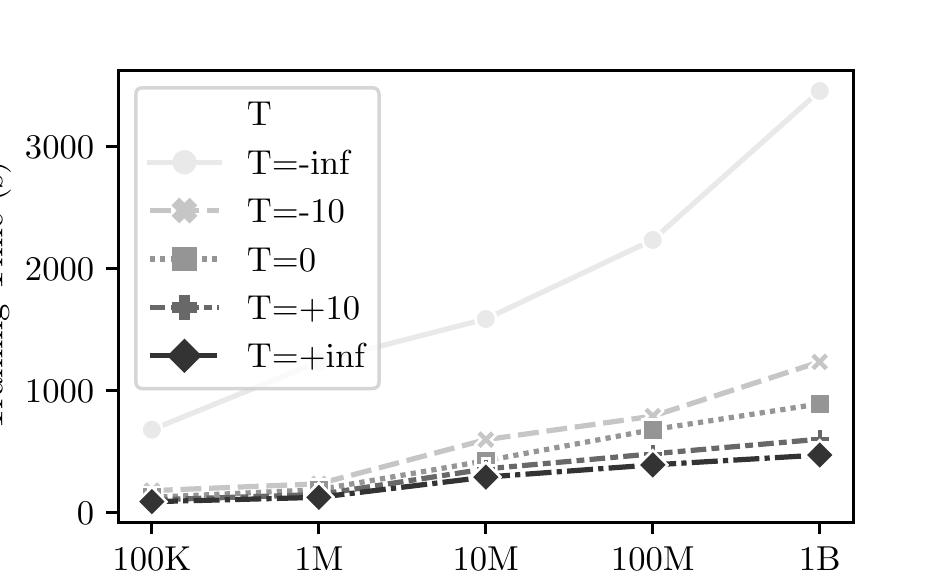}
         \caption{Performance of Model Building}
         \label{fig:training_vae}
     \end{minipage}
     \hspace{1mm}
     \begin{minipage}[t]{0.32\linewidth}
         \centering
         \includegraphics[width =\textwidth]{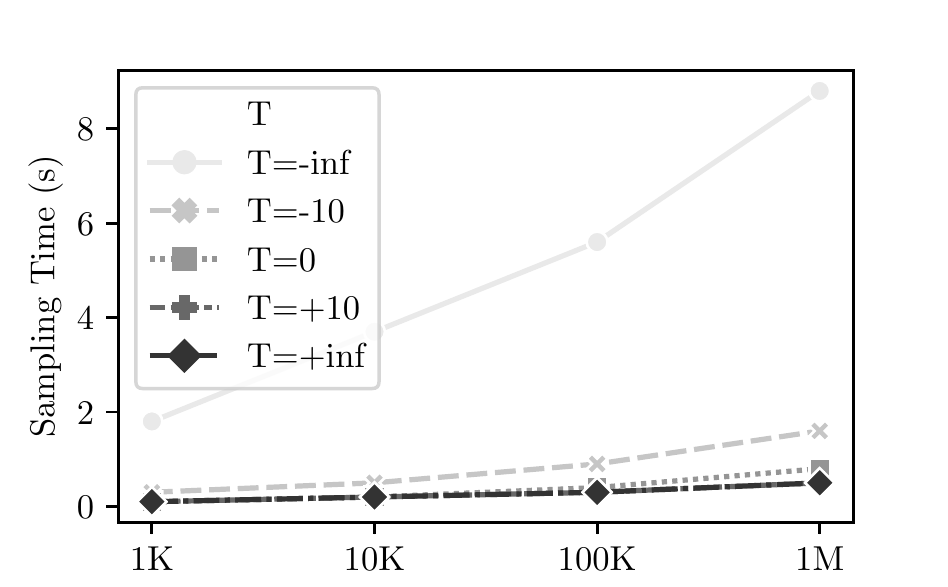}
         \caption{Performance of Sample Generation}
         \label{fig:sampling_vae}
     \end{minipage}
 \end{figure*}

\stitle{Impact of Rejection Sampling.}
Figure~\ref{fig:VAERejectionSampling} shows how varying the value of $T$ impacts the sample quality.
Recall from Section~\ref{subsec:approximationErrors} that as $T \rightarrow +\infty$, almost all samples from VAE are accepted,
while when $T \rightarrow -\infty$, samples are rejected unless they are likely to be from the true posterior distribution.
As expected, decreasing the value of $T$ results in decreased value of relative error difference.
However, this results in a larger number of samples being rejected.
Our approach allows $T$ to be varied across queries such that queries with stringent accuracy requirements can use small $T$ for better estimates.
We investigate the impact of rejection sampling on model building and sample generation later in the section.

\stitle{One versus Multiple VAEs.}
In the next set of experiments, we consider the case where one uses multiple VAEs to learn the underlying data distribution.
We partitioned the attributes based on marital-status for Census and origin-state for Flights.
We evaluated partitioning data over other attributes and observed similar results.
In order to compare the models fairly, we ensured that the cumulative model capacity for both scenarios were the same.
For example, if we built $K$ VAE models with capacity $C$ each, then we compared it against a single VAE model with capacity $K \times C$.
Figure~\ref{fig:oneVsManyVAE} shows the results.
As expected, the sample quality improves with larger number of VAE models enabling them to learn finer data characteristics.
Interestingly, we observe that increasing the model capacity for the single VAE case has diminishing returns due to the fixed size of the training data.
In other words, increasing the capacity does not improve the performance beyond certain model capacity.
Figure~\ref{fig:oneVsManyVAEDPAlgo} compares the performance of partitions selected by the dynamic programming algorithm for the scenario where an OLAP hierarchy is provided. We compare it against a greedy algorithm.
As expected, our proposed approach that is cognizant of the R-ELBO metric provides better partitions -
especially datasets such as Flight that have complex R-ELBO distributions.

\subsection{Comparison with DL Model for AQP.}
While we primarily focused on VAE, it is possible to leverage other deep generative models for AQP.
Figure~\ref{fig:allDLModels} compares the performance of three common models : VAE, GAN and Bayesian Networks (BN).
Generative Adversarial Networks (GANs)~\cite{DBLP:journals/corr/Goodfellow17,Goodfellow-et-al-2016} are a popular and powerful class of generative models that
learn the distribution as a minimax game between two components - generator (that generates data)
and discriminator (that identifies if the sample is from the true distribution or not).
(Deep) Bayesian networks (BN) are another effective generative model that specifies the joint distribution as a directed graphical model
where nodes correspond to random variable $A_i$ (for attribute $A_i$)
and directed edges between nodes signify (direct) dependencies between the corresponding attributes.
Please refer to~\cite{Goodfellow-et-al-2016} for more details.
In order to ensure a fair comparison, we imposed a constraint that the model size for all three approaches are fixed.
Furthermore, VAE provides the best results for a fixed model size.
GANs provide reasonable performance but was heavily reliant on tuning.
Training a GAN requires identifying an equilibria and tuning of many parameters such as the model architecture and learning rate  \cite{DBLP:journals/corr/Goodfellow17}. This renders the approach hard to use in practise for general data sets.
Identifying appropriate mechanisms for training GANs over relational data for AQP is a promising avenue for future research.
BNs provide the worst result among the three models.
While BNs are easy to train for datasets involving discrete attributes,
a hybrid dataset with discrete and continuous attributes,  and attributes with large domain cardinalities are challenging.
When the budget on model size is strict, BNs often learn a sub-optimal model.

We also evaluated VAE against the recently proposed MSPN~\cite{molina2018mixed} that has been utilized for AQP in~\cite{kulessa2018model}.
Similar to Bayesian Networks, MSPNs are acyclic graphs (albeit rooted graphs) with sum and product nodes as internal nodes.
Intuitively, the sum nodes split the dataset into subsets while product nodes split the attributes.
The leaf nodes define the probability distributions for an individual variable.
MSPN could be used to represent an arbitrary probability distribution~\cite{molina2018mixed}.
We used the random sampling procedure from~\cite{kulessa2018model} for generating samples from a trained MSPN.
We observed that MSPN often struggles to model distributions involving large number of attributes and/or tuples
and that using a single MSPN for the entire model did not provide good results. As a comparison to train a VAE on 1M tuples of the Census data set on all attributes requires a few minutes versus almost 3.5 hours for MSPN. In addition the accuracy of queries with larger number of attributes for the case of MSPN was very poor and not close to any of the other models.
Hence, we decided to provide {\em an advantage} to MSPN, building the model over subsets of attributes. That way we let the model focus only on specific queries and improve its accuracy.
There were around 120 distinct combination of measure and filter attributes in our query workload.
We built MSPN models for each combination of attributes, generate samples from it and evaluate our queries over it.
For example, if a query involved an aggregate over $A_x$ and filter condition over $\{A_i, A_j,A_k\}$,
we built an MSPN over the projected dataset containing only $\{A_x,A_i,A_j,A_k\}$.
Unlike GAN and BN, we did not control the number of leaf nodes.
However, the size of the MSPN models that were trained over attribute subsets were in the same ballpark as the other generative models.
Figure~\ref{fig:allDLModels} presents the performance of VAE and MSPN (build on specialized subsets of attributes) to be superior over GAN and BN.
However, in the case of VAE the model was trained over the entire dataset being able to answer {\em arbitrary queries} while MSPN was trained over specific attribute subsets utilized by specific queries.
Even in this case, providing full advantage to MSPN, the median relative error difference for VAE and MSPN were 0.060835 and 0.137699 respectively, more than two times better for VAE.
This clearly demonstrates that a VAE model can learn a better approximation of the data, being able to answer {\em arbitrary queries} while it can be trained an order of magnitude faster than MSPN as detailed next.

Next, we compare our approach with DBEst~\cite{dbest} and NeuralCubes~\cite{DBLP:journals/corr/abs-1808-08983} that use ML models for answer AQP queries.
Figure~\ref{fig:allDLModels} compares the performance of our approach against these methods.
In contrast to our approach that uses synthetic samples, DBEst and NeuralCubes use pre-built models to directly answer AQP queries.
For simple aggregate queries, the performance of both these methods are comparable to that of our approach.
However, our approach produces more accurate result for ad-hoc queries that are very common in data exploration.
Furthermore, the ability of our approach to create arbitrary number of samples to achieve low error that is not possible with DBEst and NeuralCubes.

\stitle{Performance Experiments.}
Our next set of experiments investigate the scalability of VAE for different dataset sizes and values of threshold $T$.
Figure~\ref{fig:training_vae} depicts the results for training over a single GPU. All results would be substantially better with the use of multiple GPUs.
As expected, the training time increases with larger dataset size.
However, due to batching and other memory optimizations, the increase is sublinear.
Next, incorporating rejection sampling has an impact on the training time with stringent values of $T$ requiring more training time.
The increased time is due to the larger number of training epochs needed for the model to learn the distribution.
The validation procedure for evaluating the rejection rate uses a Monte Carlo approach~\cite{grover2018variational} that also contributes to the increased training time. However overall it is evident from our results that very large data sets can be trained very efficiently even on a single GPU. This attests to the practical utility of the proposed approach.
Figure~\ref{fig:sampling_vae} presents the cost of generating samples of different sizes and for various values of $T$.
Not surprisingly, lower values of $T$ require a larger sampling time due to the higher number of rejected samples.
As $T$ becomes less stringent, sampling time dramatically decreases.
Interestingly, the sampling time does not vary a lot for different sampling sizes.
This is due to the efficient vectorized implementation of the sampling procedure in PyTorch and the availability of larger memory that could easily handle samples of large size. It is evident again that the proposed approach can generate large number of samples in fractions of a second making the approach highly suitable for fast query answering with increased accuracy.

%% file: relatedWork.tex
\section{Related Work}
\label{sec:relatedWork}

\stitle{Deep Learning for Databases.}
Recently, there has been increasing interest in applying deep learning techniques for solving fundamental problems in databases.
SageDB~\cite{kraska2019sagedb} proposes a new database architecture that integrates deep learning techniques to model
data distribution, workload and hardware and use it for indexing, join processing and query optimization.
Deep learning has also been used for learning data distribution to support index structures~\cite{kraska2018case},
join cardinality estimation~\cite{kipf2018learned,ortiz2018learning}, join order enumeration~\cite{krishnan2018learning,marcus2018deep},
physical design~\cite{pavlo2017self}, entity matching~\cite{ebraheem2018distributed},
workload management~\cite{marcus2017releasing} and performance prediction~\cite{venkataraman2016ernest}.

\stitle{Sampling based Approximate Query Processing.}
AQP has been extensively studied by the database community.
A detailed surveys is available elsewhere~\cite{garofalakis2001approximate,mozafari2015handbook}.
Non sampling based approaches involve synopses data structures such as histograms, wavelets and sketches.
They are often designed for specific types of queries and could answer them efficiently.
In our paper, we restrict ourselves to sampling based approaches~\cite{acharya1999aqua,Agarwal:2013:BQB:2465351.2465355,park2018verdictdb, kandula2016quickr,chaudhuri2007optimized}.
Samples could either be pre-computed or obtained during runtime.
Pre-computed samples often leverage prior knowledge about workloads to select samples that minimize the estimation error.
However, if workload is not available or is inaccurate, the chosen samples could result in worse approximations.
In this case, recomputing samples is often quite expensive.
Our model based approach could easily avoid this issue by generating samples as much as needed on-demand.
Online aggregation based approaches such as~\cite{hellerstein1997online,wu2010continuous}
continuously refine the aggregate estimates during query execution.
The execution can be stopped at any time if the user is satisfied with the estimate.
Prior approaches often expect the data to be retrieved in a random order which could be challenging.
Our model based approach could be easily retrofitted into online aggregation systems as they could generate random samples efficiently.
Answering ad-hoc queries and aggregates over rare sub-populations is especially challenging~\cite{Chaudhuri:2017:AQP:3035918.3056097} .
Our approach offers a promising approach where as many samples as needed could be generated to answer such challenging queries without having to access the dataset.
\cite{kulessa2018model} uses mixed sum-product networks (MSPN) to generate aggregate estimates for interactive visualizations.
While in the same spirit as our work, their proposed approach suffers from scalability issues that limits its widespread applicability.
Even for a small dataset with 1 million tuples, it requires hours for training. This renders such an approach hard to apply for very large data sets. In contrast a VAE model can be trained in a matter of minutes making it ideal for very large data sets.

%% file: conclusion.tex
\section{Conclusion}
\label{sec:conclusion}

We proposed a model based approach for AQP and demonstrated
experimentally that the generated samples are realistic and produce accurate aggregate estimates.
We identify the issue of model bias and propose a rejection sampling based approach to mitigate it.
We proposed dynamic programming based algorithms for identifying optimal partitions to train multiple generative models.
Our approach could integrated easily into AQP systems and
can satisfy arbitrary accuracy requirements by generating as many samples as needed without going back to the data.
There are a number of interesting questions to consider in the future.
Some of them include better mechanisms for generating conditional samples that satisfy certain constraints. Moreover, it would be interesting to study the applicability of generative models in other data management problems such as synthetic data generation for structured and graph databases extending ideas in \cite{DBLP:journals/pvldb/ParkMGJPK18}.